# Modelling the photochrome-TiO$_2$ interface with Bethe-Salpeter and TD-DFT methods

*Daniel Escudero,[a] Ivan Duchemin,[b,c] Xavier Blase,[c] and Denis Jacquemin [a,d,*]*

[a]CEISAM UMR CNRS 6230, Université de Nantes, 2 rue de la Houssinière, BP 92208, 44322 Nantes Cedex 3, France. E-mail : Denis.Jacquemin@univ-nantes.fr

[b]INAC, SP2M/L\Sim, CEA/UJF Cedex 09, 38054 Grenoble, France

[c]Univ. Grenoble Alpes, CNRS, Inst NEEL, F-38042 Grenoble, France

[d]Institut Universitaire de France, 1 rue Descartes, F-75005 Paris Cedex 05, France

KEYWORDS: TD-DFT; BSE/*GW*; exchange-correlation functionals; photochromes; azobenzene; hybrid organic/inorganic systems; interface.

**Abstract:** Hybrid organic/inorganic-semiconductor systems have important applications in both molecular electronics and in photo-responsive materials. The characterization of the interface and of the electronic excited-states of these hybrid systems remains a challenge for state-of-the-art computational methods, as the systems of interest are large. In the present investigation, we present for the first time a many-body Green's function Bethe-Salpeter investigation of a series of photochromic molecules adsorbed onto TiO$_2$ nanoclusters. Based on these studies, the performance of TD-DFT is assessed. Using a state-of-the-art computational protocol, the photochromic properties of different hybrid systems are assessed. This work shows that qualitatively different conclusions can be reached with TD-DFT relying on various exchange-correlation functionals for such organic/inorganic interfaces, and paves the way to more accurate simulation of many materials.

The functionalization of semiconductor surfaces with photo-responsive organic and inorganic materials has led to multiple applications in molecular electronics,[1-2] including dye-sensitised solar cells (DSSCs)[3] and photosensor/photoswitching materials.[2,4] First, functional photochromes grafted onto semiconductor hybrid systems may act as sensors of the molecular switch state, since changes in the conductance or in the open circuit voltage usually accompany the light-induced photoconversion. These systems have been largely studied, mainly from an experimental viewpoint,[5-6] in an effort to achieve a control of the photochromic properties of the joint systems. Importantly, upon grafting the photochrome unit onto the semiconductor surface, its spectroscopic, electronic and photochromic properties might be modulated or completely altered (leading for instance to an electron injection process), although systems where the photochromic properties are successfully retained have also been reported.[7-8,9] This illustrates the interplay between the photoswitches and the underlying surface. First-principles modeling of these hybrid systems helps rationalizing and predicting a putative photochromic behavior, and our group has performed such studies for azobenzene (AZB) and dithyenylethene (DTE)-$TiO_2$ hybrid systems,[10-11] using periodic slabs to model the interface. Due to the large size of these hybrid systems, their computational modelling is still restricted, in practice, to density functional theory (DFT) and time-dependent DFT (TD-DFT) investigations. However, the performance of these methods for these systems is strongly functional-dependent and the treatment of the optical and the semiconductor gaps on an equal footing remains a challenge for these methods. As potential alternatives to tackle these systems, the many-body Green's function Bethe-Salpeter equation (BSE) formalism,[12,13,14] an excited-state method that builds upon the many-body *GW* calculation of occupied and virtual energy levels,[15,16,17] appears as a promising candidate. BSE/*GW* was initially developed for extended periodic semiconductors, but recent Gaussian-based implementations have allowed the approach to become recently popular for gas phase organic and hybrid systems.[18,19,20,21] Indeed, as recently shown for the Thiel's benchmark set of organic molecules,[22] the BSE/*GW* formalism can outperform adiabatic TD-DFT for both excitation energies[23] and oscillator strengths.[24] Importantly when a self-consistent BSE/ev*GW* approach is applied, namely an approach where the underlying occupied/virtual *GW* electronic energy levels are self-consistently converged, the strong dependency on the functional obtained with both TD-DFT and the perturbative BSE/$G_0W_0$ scheme[25] is almost completely washed out. It was further demonstrated that the BSE formalism is adequate to accurately intra[26,27] and intermolecular[28,29] charge-transfer (CT) excitations due to the non-locality of the screened Coulomb potential operator that couples non-overlapping electron and hole distributions. In this contribution, we assess for the first time the performance of the BSE/*GW* method for the spectroscopic, electronic and photochromic properties of the photochrome-$TiO_2$ interface. Furthermore, based on the BSE/*GW* results, the performance of TD-DFT methods is further evaluated. We anticipate that the conclusions extracted from this study can be extrapolated to other hybrid systems, for instance to modelling the dye-$TiO_2$ interface in DSSCs.

For the description of the chromophore-$TiO_2$ interface we use an anatase cluster model, which exposes the majority (101) surfaces. More in details, we use two stoichiometric $(TiO_2)_n$ clusters, with n=38 and 25. These clusters possess all possible oxygen positions doubly saturated and they are globally neutrally charged. Such cluster models are able to nicely reproduce the density of states (DOS) of periodic $TiO_2$ surfaces.[30] As usual, we considered a bidentate anchoring of the carboxylic group onto the $TiO_2$, see Figure 1, where the acidic proton atom is transferred to an oxygen atom of the surface. Geometry optimizations were performed at the CAM-B3LYP/6-31G(d) level of theory. These optimization led to a global orientation of the photochromes onto the surface similar to the one obtained with periodic calculations for the same system.[10] The choice of CAM-B3LYP/6-31G(d) for determining the geometries is also justified by the fact that this approach provides a N=N bond length of 1.247 Å for the isolated molecule in its most-stable form, in perfect agreement with the 1983 X-ray determination of Bouwstra and co-workers.[31] Vertical excitation energies were obtained with TD-DFT using a very popular hybrid functionals, namely, PBE0, as well as two widely used range-separated hybrid functionals, i.e., CAM-B3LYP and ωB97X-D functionals at the optimized geometries using the same basis set as in the optimizations. All DFT/TD-DFT calculations were made with the Gaussian09 code.[32] Our *GW* and Bethe-Salpeter equation (BSE) calculations were performed with the FIESTA package[29,33,34] that uses Gaussian bases combined with the Coulomb-fitting resolution-of-identity approach. For sake of comparison with TD-DFT results, the BSE/*GW* calculations were performed with the same 6-31G(d) basis and the Weigend Coulomb Fitting auxiliary basis.[35] As stated above, we adopt the ev*GW* approach,[36] starting from Kohn-Sham eigenstates generated with the M06-2X functional[37] as implemented in the NWChem package.[38] Such a combination has been shown to be optimal elsewhere for organic compounds.[39]

The chromophores **1**-**6** studied herein are displayed in Scheme 1. This study covers three different functionalized AZB molecules in their *trans* (**1**,**3**,**5**) and *cis* (**2**,**4**,**6**) forms. Their photoconversion mechanisms are also highlighted in Scheme 1. All the derivatives bear a terminal –COOH group, so that they can be anchored onto $TiO_2$. To model the photochrome-$TiO_2$ interface, two different $TiO_2$ cluster models were employed, i.e., **$(TiO_2)_{38}$** and **$(TiO_2)_{25}$**. The larger cluster has been previously used to model the dye-TiO$_2$ interface in DSSCs[40] and has been proved successful as compared to periodic models.[41] The smaller cluster was used for more computational demanding BSE/*GW* calculations and these results were used as benchmarks to assess the TD-DFT's performances.

**Figure 1. a)** Bidentate anchoring of the carboxylate⁻ group to the TiO$_2$ slab; **b)** cenital view of the chromophore-TiO$_2$ interface; **c)** frontal view of the *cis*-azobenzene-TiO$_2$ interface highlighting the acidic proton atom, which is transferred from the photochrome. H, C, N, O and Ti atoms are in white, brown, light blue, red and orange colors, respectively.

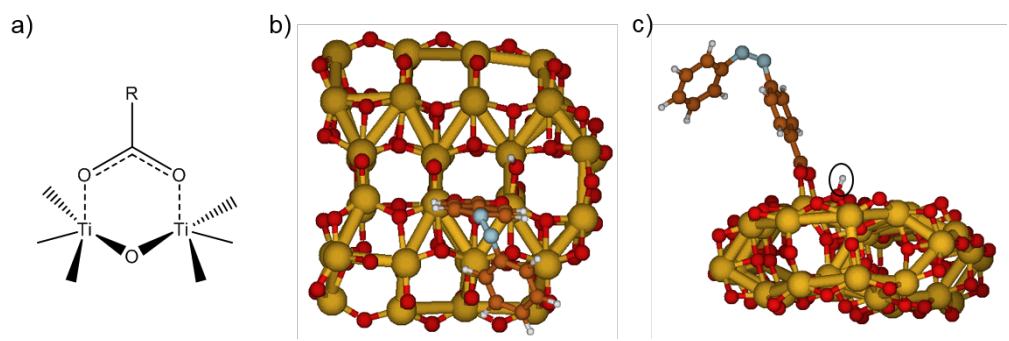

**Scheme 1. Chemical structure of chromophores 1-6. Odd (even) numbers correspond to the *trans* (*cis*) isomers.**

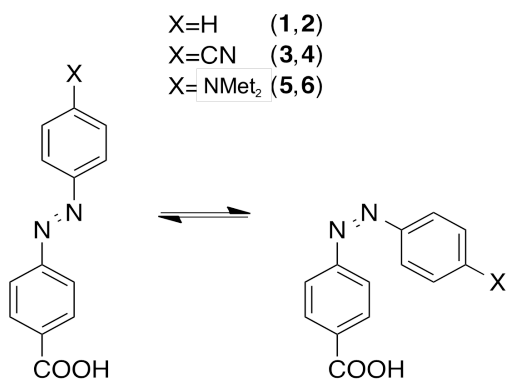

Table 1 lists the TD-DFT and BSE/ev$GW$ results for the **1,2-(TiO$_2$)$_{25}$** and **1,2-(TiO$_2$)$_{38}$** interface systems. The low-energy UV-Vis spectrum of the isolated *trans*- and *cis*-AZB is dominated by $n\pi^*$ (S$_1$) and $\pi\pi^*$ (S$_2$) excitations, and the photochromic properties of AZBs are known to be associated with the population of these two excited-states.[42] First, we explore the character of the low-lying excited-states of AZBs upon grafting them on a TiO$_2$ surface. The results for the low-lying states are quite similar regardless of the selected cluster model used, i.e., the results for **1-(TiO$_2$)$_{25}$** and **1-(TiO$_2$)$_{38}$** are similar. According to the best theoretical estimates, that is the BSE/ev$GW$ calculations, S$_1$ and S$_2$ are the photoactive $n\pi^*$ and $\pi\pi^*$ excited-states. Therefore, the natures of the low-lying excited-states of these hybrid systems are unchanged compared to the ones obtained at the gas-phase or solution and one can infer that the photoswitching properties of AZBs will be preserved on the anatase surface. These computational evidences are in agreement with the observed photoswitching abilities of related AZBs-TiO$_2$ systems.[7] The comparison of the BSE and TD-DFT results reveals that only the two range-separated functionals, i.e., CAM-B3LYP and ω97X-D, are able to accurately restore both the excited-states energies (within ca. 0.15 eV of the BSE/ev$GW$ results) and oscillator strengths (for $n\pi^*$ and $\pi\pi^*$ states of **1-TiO$_2$** and for the $n\pi^*$ state of **2-TiO$_2$**, see Table 1). Additionally, these functionals provide the correct state ordering in most of the cases. This is in sharp contrast with the results obtained with the hybrid functional, PBE0, which considerably underestimates the excitation energies of both states and significantly changes the corresponding state order. Thus, while the $n\pi^*$ state remains S$_1$, the $\pi\pi^*$ state is found as an upper-lying state, e.g., S$_4$, in the **1-(TiO$_2$)$_{38}$** model. In between $n\pi^*$ and $\pi\pi^*$ states, spurious CT states are found with PBE0, as a result of the spurious redshift of CT excitations when semilocal functionals with limited amount of *exact* exchange are used. While the appearance of such spurious states is not uncommon for systems with strong CT nature, this is one of the first examples in which an accurate functional for gas phase calculations leads to erroneous conclusions upon functionalization onto a surface. This behavior could be expected as the surface plays the role of an acceptor, therefore changing the underlying photophysics. In this framework, we recall that global hybrids like B3LYP and PBE0 have been often used to model dyes grafted onto TiO$_2$ for solar cell applications, and that they did provided results in agreement with experimental outcomes,[30,43,44] hinting that the exact nature of the system under investigation is of importance. In our case, this conclusion might have a large impact on the interpretation. Here, the problem is particularly remarkable for the *cis* system, where the first bright TD-PBE0 state with a certain $\pi\pi^*$ character is the 109$^{th}$ root for **2-(TiO$_2$)$_{25}$**, see Table 1. Figure 2 provides the electronic density difference (EDD) plots of the $n\pi^*$ and $\pi\pi^*$ states for **1-(TiO$_2$)$_{38}$** as obtained by different levels of theory. The $\pi\pi^*$ state at the TD-PBE0 level of theory has some degree of CT character to the TiO$_2$ moiety, while TD-CAM-B3LYP provides a state fully localized on the AZB. The better quality of the TD-CAM-B3LYP picture for both states is further confirmed by the plot of the electron-averaged hole density (white), and hole-averaged electron density (pink), associated with their corresponding BSE $\psi_\lambda(r_e, r_h)$ eigenstates given in Figure S1 (see Supporting Information). As such, standard global hybrids not only fail in the description of the spectroscopic states (energies and oscillator strengths) of the hybrid organic/inorganic systems, but the qualitative picture they provide for the photoswitching properties are incorrect. Indeed, with these functionals one would wrongly conclude that charge injection takes place rather than photochromism. There are a few experimental works treating azo switches grafted onto TiO$_2$,[7,45,46] but only one is directly comparable to our model,[46] as the others use longer linkers.[7,45] In Ref. 46, it is indeed found that the azo dye grafted onto a TiO$_2$ nanoparticle through a carboxylic group shows photochromism, consistent with the theoretical findings of BSE/ev*GW* and TD-CAM-B3LYP. Therefore, the spectroscopic properties of compounds **3-6** have been studied at the TD-CAM-B3LYP level.

**Table 1.** TD-DFT and BSE/GW excitation energies (in eV) and oscillator strengths (between parentheses) of the low-lying *n*π* and ππ* excited-states of the hybrid systems. Note that the ππ* transitions are not always the S2 excitations, especially with TD-PBE0.

| System | State | TD-CAM-B3LYP | TD-ωB97X-D | TD-PBE0 | BSE/ev*GW* |
|---|---|---|---|---|---|
| 1-(TiO$_2$)$_{25}$ | nπ$^*$ | S$_1$: 2.57 (0.003) | S$_1$: 2.54 (0.002) | S$_1$: 2.40 (0.003) | S$_1$: 2.62 (0.002) |
| | ππ$^*$ | S$_2$: 3.76 (1.536) | S$_2$: 3.79 (1.536) | S$_{20}$: 3.42 (0.357) | S$_2$: 3.91 (1.561) |
| 1-(TiO$_2$)$_{38}$ | nπ$^*$ | S$_1$: 2.64 (0.000) | S$_1$: 2.62 (0.000) | S$_1$: 2.44 (0.000) | |
| | ππ$^*$ | S$_2$: 3.91 (1.563) | S$_2$: 3.96 (1.580) | S$_4$: 3.44 (0.376) | |
| 2-(TiO$_2$)$_{25}$ | nπ$^*$ | S$_1$: 2.55 (0.055) | S$_1$: 2.56 (0.054) | S$_2$: 2.46 (0.079) | S$_1$: 2.62 (0.062) |
| | ππ$^*$ | S$_4$: 4.25 (0.335) | S$_2$: 4.36 (0.529) | S$_{109}$: 4.29 (0.061) | S$_2$/S$_3$: 4.36 (0.136)/4.50 (0.272) |
| 2-(TiO$_2$)$_{38}$ | nπ$^*$ | S$_1$: 2.72 (0.048) | S$_1$: 2.73 (0.046) | S$_1$: 2.61 (0.081) | |
| | ππ$^*$ | S$_2$: 4.29 (0.414) | S$_3$: 4.45 (0.594) | S$_{57}$: 4.15 (0.041) | |

**Figure 2.** Representation of the TD-DFT density difference plots of the nπ$^*$ and ππ$^*$ states of 1-(TiO$_2$)$_{38}$.

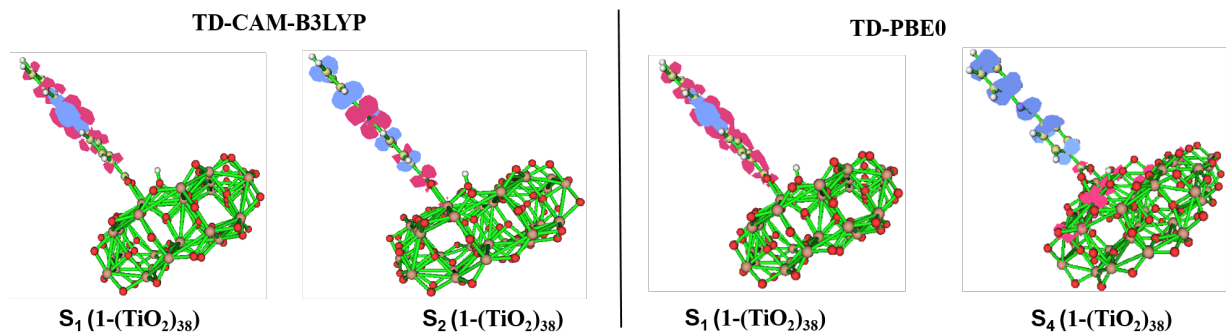

The TD-CAM-B3LYP results for the hybrid organic/inorganic systems **3-6** are presented in Table 2. Within this series, the effect of substituting AZB with electron-withdrawing (**3-4**) and electron-donor (**5-6**) moieties is assessed. While the $n\pi^*$ state is merely unaffected by the substitution pattern (and thereto the photoswitching properties associated to the $n\pi^*$ state), the $\pi\pi^*$ state is importantly modulated by substitution. Hence, the cyano substitution leads to bluer bands than its amino counterpart. Brighter $\pi\pi^*$ bands are obtained with the former substitution pattern. Furthermore, the $\pi\pi^*$ state remains a low-lying state ($S_2$ and $S_3$) for the cyano compounds. In view of all pieces of evidence, the electron-withdrawing substitution of AZB seems to enhance the photoswitching properties associated to the $\pi\pi^*$ state. This follows chemical intuition, as the cyano group will withdraw density far from the surface whereas the donor amino group will push the density towards the surface, which might be detrimental for the photochromic activity.

**Table 2.** TD-CAM-B3LYP (in eV) and oscillator strengths (between parentheses) of the n$\pi$* and $\pi\pi$* excited states of the hybrid systems 3-6. Note that the $\pi\pi$* transitions are not always the S2 excitations.

| | System | 3-(TiO$_2$)$_{38}$ | 4-(TiO$_2$)$_{38}$ | 5-(TiO$_2$)$_{38}$ | 6-(TiO$_2$)$_{38}$ |
|---|---|---|---|---|---|
| State | $n\pi^*$ | S$_1$: 2.61 (0.000) | S$_1$: 2.70 (0.049) | S$_1$: 2.70 (0.000) | S$_1$: 2.75 (0.066) |
| | $\pi\pi^*$ | S$_2$: 3.89 (1.791) | S$_3$: 4.47 (0.561) | S$_{20}$: 3.25 (1.677) | S$_{34}$: 4.24 (0.156) |

In conclusions, the interface, spectroscopic and photochromic properties of hybrid organic/inorganic-semiconductor systems are assessed with the GW/*BSE* and TD-DFT formalisms. More specifically, several AZB photochromes adsorbed onto $TiO_2$ nanoclusters were studied. The BSE calculations relied on partially self-consistent ev*GW* electronic occupied/virtual energy levels so as to limit the dependency on the starting functional and obtain accurate theoretical estimators. These estimates have been used to benchmark the TD-DFT results. Among the different exchange-correlation functionals used in the TD-DFT calculations, only range-separated functionals are able to accurately describe the excited-states of these hybrid organic/inorganic-semiconductor systems, and standard global hybrids overshoot charge injection though these functionals might provide an accurate description of the isolated photochrome. This comes has a warning for calculations of electroactive devices, e.g., DSSCs modeling of organic/inorganic interfaces with non-optimal theoretical approaches. Different substitution patterns of the AZB photocrome were studied with a range-separated functional and the obtained results indicate that the electron-withdrawing substitution of AZB generally enhances the photoswitching capabilities at the hybrid systems.

AUTHOR INFORMATION

**Corresponding Author**

*Denis.Jacquemin@univ-nantes.fr

ACKNOWLEDGMENTS

D.E. acknowledges the European Research Council (ERC) and the *Région des Pays de La Loire* for his post-doctoral grant (Marches – 278845) and the European Union's Horizon 2020 research and innovation programme under the Marie Sklodowska-Curie grant agreement No 700961. D.J.

thanks the ERC for his support in the framework of the Marches project (n° 278845). This research used resources from the GENCI French national supercomputing resources and of the CCIPL.

SUPPORTING INFORMATION

BSE electron-hole plot and the Cartesian coordinates of all complexes are presented in the Supporting Information.